\documentclass[aps,prd,preprint,floatfix,preprintnumbers,nofootinbib,superscriptaddress]{revtex4}
\usepackage[utf8]{inputenc}
\usepackage{fancybox}
\usepackage{hhline}
\usepackage{dcolumn}
\usepackage{textcomp}
\usepackage{epsfig,graphics,graphicx}
\usepackage{amsfonts,amssymb,amsmath}
\usepackage{pifont}
\usepackage{bm}
\usepackage{longtable} 
\usepackage{appendix}
\usepackage{lscape}
\usepackage[mathscr]{euscript}
\usepackage{mathrsfs}
\usepackage{multirow}
\usepackage{rotating}
\usepackage{color}

\newcommand{\beq}{\begin{equation}}
\newcommand{\eeq}{\end{equation}}
\newcommand{\bea}{\begin{eqnarray}}
\newcommand{\eea}{\end{eqnarray}}

\newcommand{\eps}{\epsilon}
\newcommand{\ord}[1]{{\cal{O}}( #1 )}
\newcommand{\B}{{\bf B}}
\newcommand{\Bdag}{{\bf B^\dagger}}
\newcommand{\p}{{\mathfrak{p}}_0}

\usepackage{sansmath}
\usepackage{lmodern}
\newcommand{\gA}{\mathring g_A}
\newcommand{\BChPTNc}{$BChPT\times{ 1/N_c}~$}

\DeclareFontFamily{OT1}{pzc}{}
\DeclareFontShape{OT1}{pzc}{m}{it}%
              {<-> s * [0.900] pzcmi7t}{}
\DeclareMathAlphabet{\mathpzc}{OT1}{pzc}%
                                 {m}{it}
\DeclareMathAlphabet{\mathcalligra}{T1}{calligra}{m}{n}
\usepackage{hyperref}
\hypersetup{pdfauthor={me}, colorlinks=true, citecolor=blue, urlcolor=blue, linkcolor=black}
\begin{document}
\preprint{\vbox{\hbox{ JLAB-THY-19-2916} }}
\title{\phantom{x}
\vspace{-0.5cm}     }
\title{\centering  BChPT$\times$1/N${  \rm  \sansmath_c}$  in SU(3): a more effective theory  }
\author{Ishara P. Fernando}

\affiliation{	Hampton University \& Thomas Nelson Community College,  Hampton , VA, USA}
\email{ishara@jlab.org}
\author{Jos\'{e} L. Goity}
\thanks{Speaker.  9th International workshop on Chiral Dynamics,
		17-21 September 2018, 
		Durham, NC, USA }
\affiliation{	Hampton University,  Hampton , VA  \& Jefferson Lab, Newport News, VA, USA}
\email{goity@jlab.org}
\begin{abstract} The   chiral and  $1/N_c$ expansions are combined in the description of low energy baryons. The combination  furnishes a better behaved expansion, consequence of exactly eliminating large terms   that violate the $1/N_c$ expansion and which are typical of  formulations of BChPT where  consistency with the large $N_c$ limit  is disregarded. The  improvements are particularly dramatic in the case of SU(3).   The general framework is outlined and applications to the vector charges and axial couplings are presented along with  a comparison with  Lattice QCD results  with three quark flavors. 
\end{abstract}
\maketitle
\section{Introduction}
 Key baryon observables depend on the number of colors $N_c$ of QCD,  most notably  the baryon masses and the axial-vector couplings which are   $\ord{N_c}$. Since the latter determine the strength of the couplings of Goldstone Bosons (GB) to baryons ($g_{\pi N}$, etc.),  consistency with  the large $N_c$ limit imposes  fundamental  constraints on the low energy  effective theory.  Such constraints result from the observation that a consistent theory of baryons at large $N_c$ must enforce a dynamical spin-flavor symmetry \cite{Gervais:1983wq,Gervais:1984rc,Dashen:1993as,Dashen:1993ac}. Such a symmetry is non-relativistic in nature, consistent with the fact that in large $N_c$  baryons are heavy, and has the form of a contracted $SU(2N_f)$  symmetry,  broken by effects $\ord{1/N_c}$. The spin-flavor symmetry requires that the baryon states furnish $SU(2N_f)$ irreducible representations (irrep) with (at least) $N_c$ boxes in the Young tableux. In particular, ground state baryons correspond to the totally symmetric irrep with $N_c$ boxes, which spans baryon spins from $S=1/2$ to $N_c/2$ ($N_c$ assumed odd). The generators of the spin-flavor symmetry are the baryon spin $S^i$, the flavor $T^a$ and the spin-flavor $G^{ia}$ operators, where the latter are  at  leading order in $1/N_c$ proportional to the spatial components of the axial currents.  In particular, the matrix elements of the generators $G^{ia}$ are $\ord{N_c}$ and reflect the fact that baryon axial currents are of that order.  The spin-flavor symmetry is a large $N_c$ symmetry of the baryon spectrum, and constrains the couplings of different operators to baryons, e.g., axial and vector currents, scalar currents, etc., which are represented at the baryon level as composite operator products of the $SU(2N_f)$  generators. On the other hand the spin-flavor symmetry is not a Noether symmetry, and thus it only leads to constraints on the  couplings (LECs) of the effective theory. In order to implement those constraints in BChPT, one proceeds as follows. The ground state baryons are represented by a baryon field \B~ which is in the symmetric irrep of $SU(2N_f)$ and  consists  of a tower   to spins  ranging from $S=1/2$ to $N_c/2$. For $N_f=2$ the baryon isospin is equal to the spin, and for $N_f=3$ the $SU(3)$ irrep of each baryon in the tower is determined by the spin and given by $(p,q)=(2 S, \frac 12 (N_c-2 S))$. The chiral transformations are then implemented as  usual for matter fields (see \cite{CalleCordon:2012xz} for details).  Since baryons have mass $\ord{N_c}$, it is natural to use HBChPT  as the expansion in powers of $1/m$ becomes part of the $1/N_c$ expansion.  The mass splitting between baryons of different spin (hyperfine splitting) is driven by effects $\ord{1/N_c}$, so for instance the $\Delta$-$N$ mass difference is of that order. This becomes a small mass scale in addition to the GB masses, both entering together in non-analytic terms of the low energy expansion. For this reason the chiral and $1/N_c$ expansions are not independent and must be linked by choosing the relative order of the GB masses and the hyperfine splitting. The natural choice that works best in the real world with $N_c=3$ is that in which non-analytic terms are not expanded, thus corresponding to $\ord{1/N_c}=\ord{p}=\ord{\xi}$: the $\xi$ expansion, which serves to organize the effective chiral Lagrangian. The $1/N_c$ power counting is implemented in terms of the n-bodyness of composite operators:  irreducible \footnote{Irreducible means that the operator cannot be further reduced using commutation relations between its factors  or operator identities valid for the symmetric irrep of $SU(2N_f)$.} n-body operators consisting of n factors of the spin-flavor generators carry an overall factor $1/N_c^{n-1}$. 
 
 The Baryon chiral Lagrangian terms have the general structure: $\B^{\dagger} \mathbf {\Lambda_{SF}} \otimes{ \mathbf \Lambda_\chi}\; \B$, where $\mathbf {\Lambda_{SF}}$ is a spin-flavor tensor operator and $\mathbf{\Lambda_\chi}$ is a chiral tensor operator built from derivatives, sources and GB fields (see \cite{Fernando:2017yqd} for details).
The $\ord{\xi}$ Lagrangian is given by \cite{Jenkins:1995gc,CalleCordon:2012xz}:
\beq
{\cal{L}}_B^{(1)}=\Bdag (i D_0-\frac{C_{HF}}{N_c} \hat S^2-\gA u^{ia} G^{ia}+\frac{c_1}{2\Lambda}\hat \chi_+)\B
\eeq
where the term proportional to $C_{HF}$ gives the hyperfine mass splittings, $\gA=\frac{6}{5} g^N_A$ is the LO axial coupling where $g_A^N=1.2724(23)$, and the term $c_1$ gives the  $\ord{ m_q N_c}=\ord{\xi}$ baryon mass shift; $\Lambda$ is an arbitrary scale. 
It is straightforward to determine the $\xi$ power counting for loop diagrams in general. In particular the analytic pieces of the diagrams, which in particular involve the UV divergencies, can be strictly organized by powers of $p$ and $1/N_c$. In general the diagrams contributing to a given observable that has  a given power behavior in $N_c$ should not violate it. This is however not the case with the contributions of individual diagrams, but the total sum of the contributing diagrams will give the consistent result. As shown below, this can be explicitly shown  in the case of  the polynomial contributions of the different diagrams which add up to structures involving  multiple commutators of the spin-flavor generators or products thereof, restoring in this way the required $N_c$ power counting. This will be illustrated with the examples presented in this note.

As the most basic illustrative example, consider the Baryon self energy in $SU(3)$.  The polynomial contribution by the one-loop self-energy is \cite{Fernando:2017yqd}:
\bea
\delta\Sigma_{1-loop}^{\rm poly}
&=&\frac{1}{(4 \pi)^2}
\left(\frac{\mathring{g}_A
}{F_\pi}\right)^2
\left(\frac{1}{2} (\lambda_\eps+\frac 73)M_a^2[[\delta \hat m,G^{ia}],G^{ia}]-\frac{1}{3} (\lambda_\eps+\frac 83)[[\delta \hat m,[\delta \hat m,[\delta \hat m,G^{ia}]]],G^{ia}]
 \right. \nonumber\\
&+&\left. \p\;( (\lambda_\eps+1)   M_a^2 G^{ia}G^{ia}- (\lambda_\eps+2)[[\delta \hat m,[\delta \hat m,G^{ia}]],G^{ia}])\right), 
\label{eq:UVmass}
\eea
 where $F_\pi=\ord{\sqrt{N_c}}$, $\gA=\ord{N_c^0}$, $\lambda_\eps\equiv 1/\eps-\gamma+\log 4\pi$, $M_a$ are the GB octet masses,  $\delta \hat m\equiv \frac{C_{HF}}{N_c} \hat S^2$ are the hyperfine mass shifts, and $\p$ is the residual energy defined by the energy flowing through the self-energy minus the hyperfine mass shift of the corresponding state.  
One can check the large $N_c$ limit by additional expansion of the non-analytic contributions, for which one will also have similar observations concerning the $N_c$ powers as from the polynomial pieces. In the case of ordinary BChPT with only octet baryons the one loop correction to the baryon masses are finite and $\ord{m_q^{\frac 32}}$; this is not the case here where the presence of the decuplet adds non-analytic contributions which can be summarized as follows:
 i)      spin-flavor singlet piece   $\ord{1/N_c^2}$ proportional to $C_{HF}^3/F_\pi^2$; ii) spin flavor singlet $\ord{N_c^0 m_q}$ proportional to $C_{HF} m_q/F_\pi^2$;  iii)  SU(3) breaking contributions  $\ord{N_c m_q^{\frac 32}}$;  iv) quark mass independent contributions to the hyperfine mass difference $\ord{1/N_c^2}$ and quark mass dependent ones $\ord{m_q^{\frac 12}/N_c;\, m_q/N_c^2; \, m_q^{\frac 32}/N_c}$.  A key observation  is the   $\ord {m_q N_c}$ behavior of the WF renormalization constant, which  emphasizes the non-commutativity between the chiral and $1/N_c$ expansions. It will be shown that those terms play a central  role in restoring the $N_c$ power counting in one loop corrections to the various current operators. A more detailed discussion of baryon masses and the $\sigma$ terms is presented in these proceedings \cite{FernandoGoityCD18-2}. For details on the higher order Lagrangians and the results for the observables discussed here to NNLO in the $\xi$ expansion   and  generic $N_c$ see Ref. \cite{Fernando:2017yqd}.

\section{SU(3) breaking in the vector current charges }

The vector current charges are of particular interest. $SU(3)$ breaking first manifests itself through non-analytic terms in the quark masses. The Ademollo-Gatto theorem asserts that analytic terms first appear at second order in the quark masses, i.e., $\ord{p^4}$. The one loop corrections are thus finite, with all polynomial terms canceling within each of the two sets $A$ and $B$ of diagrams shown in Figure 1.

\begin{figure}[]
\begin{center}
\includegraphics[height=5cm,width=12cm,angle=0]{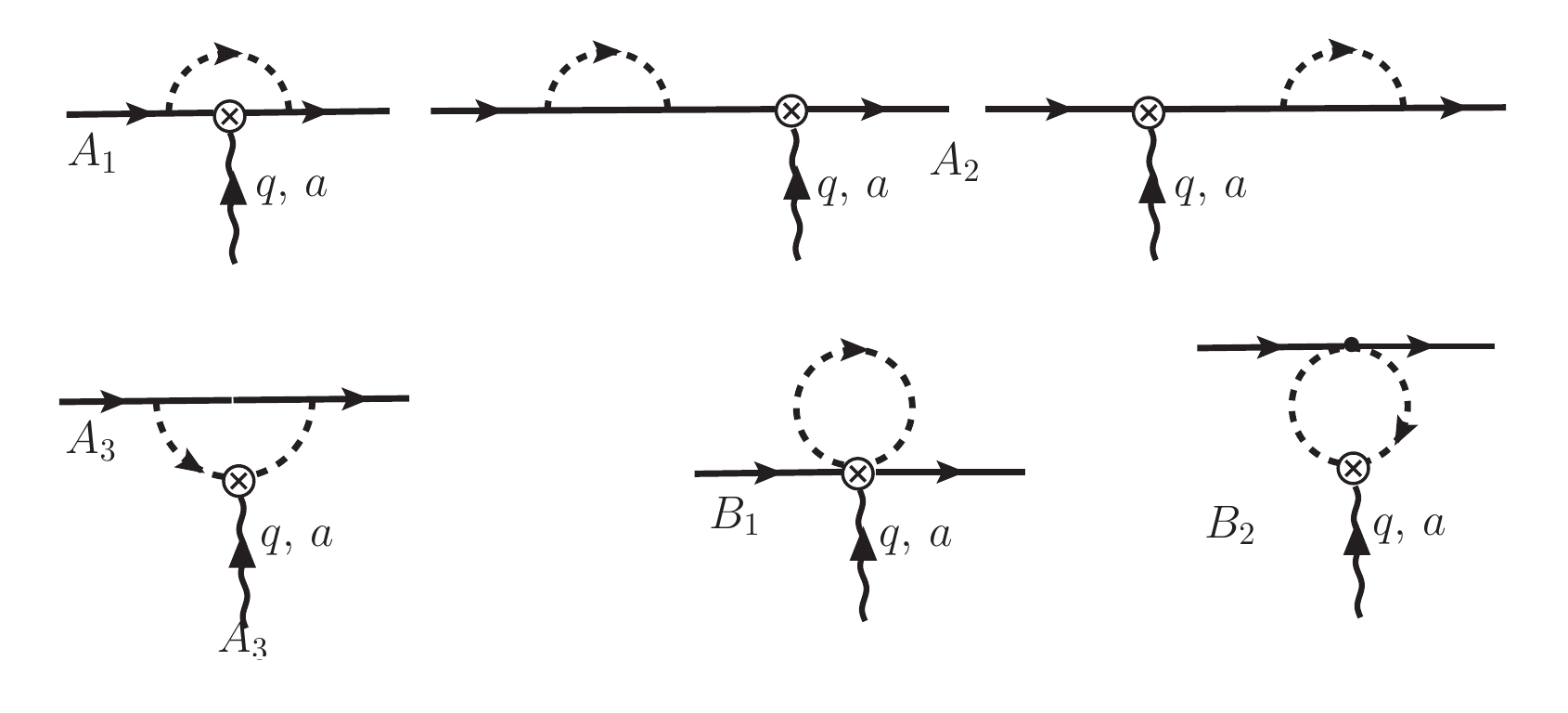}
\end{center}
\caption{One loop corrections to vector current charges.}
\label{fig:1}
\end{figure}

 As illustration of the cancellations of terms that lead to a consistent $N_c$ power counting, one can consider first the set of diagram $A_1+ A_2$. Each diagram violates $N_c$ power counting; explicit evaluation gives for the polynomial contributions \cite{Fernando:2017yqd}:
\bea
A_{1+2}^{\rm poly}&=&\frac{1}{(4\pi)^2}    \left( \frac{\mathring g_A}{F_\pi} \right)^2 \left( \frac 12 (\lambda_\eps+1) M_{ab}^2[G^{ia},[G^{ib},\Gamma]]\right.\nonumber\\
&+& \frac 13\left.(\lambda_\eps+2) \left(2[[G^{ia},\Gamma],[\delta\hat m,[\delta\hat m,G^{ia}]]]+[[\Gamma,[\delta\hat m,G^{ia}]],[\delta\hat m,G^{ia}]]\right)\right)
\eea
where $M_{ab}^2$ is the $8\times 8$ matrix representing the square of the GB masses. $\Gamma$ is any spin-flavor operator, in the present case a flavor generator $T$.  The general structure involving multiple commutators leads to the   cancellation of the $N_c$ power violating terms contributed by the individual diagrams. The key role of the wave function renormalization contribution is here evident. Explicit calculation of diagram $A_3$ shows that for the vector charge operator $A_3^{\rm poly}=-A_{1+2}^{\rm poly}$ fulfilling the AGTh. Diagrams $B$ respect $N_c$ power counting to start with.  The   non-trivial polynomial terms proportional to $q^2$ are from diagrams $A_3$ and $B_2$, contributing  to the charge radii. A complete analysis of the vector current form factors will be presented elsewhere \cite{FernandoGoity3}.
A detailed analysis of the corrections to the SU(3) charges was presented in Ref. \cite{Flores-Mendieta:2014vaa}. The consistency with the $1/N_c$ expansion can only be kept when the virtual baryons in the loop include all those that are connected via the axial current to the external baryons. This has a significant effect in the results as it had been noticed earlier when the decuplet is included in the calculations \cite{Villadoro:2006nj,Geng:2009ik,Geng:2009ik}.  The SU(3) breaking corrections to  the $\Delta S=1$ charges are therefore calculable, with the main uncertainty due to the value of  $\gA$ used, which is discussed in next section. Using the physical value  \cite{Flores-Mendieta:2014vaa}  gives results which can now be tested with LQCD calculations (as it is well known the accuracy of hyperon $\beta$ decay is not enough to pin down those corrections). Using the definitions in  \cite{Flores-Mendieta:2014vaa}, the comparison with a recent LQCD calculation \cite{Shanahan:2015dka} are satisfactory, as shown in Figure 2. Although at this point those corrections do not play a significant role in phenomenology of hyperon decays, with the use of increasingly accurate LQCD results, they will provide a significant test of the  effective theory.

\begin{table}
{\small
\begin{tabular}{lccccc}
\hline\hline\\ [-.3cm]
Charge ~~~ &~$\displaystyle \frac{f_1}{f_1^{SU(3)}}$~&&~$\displaystyle \frac{f_1}{f_1^{SU(3)}}-1~$ & ~~~~~~~~~~~~~~ & \\
 \multicolumn{2}{c}{}& \tiny{\bf [Flores-Mendieta \& Goity]}  & \tiny{\bf [Villadoro]} & \tiny{\bf [Lacour et al]} & \tiny{[Geng et al]} \\
 
  \multicolumn{2}{c}{}& \cite{Flores-Mendieta:2014vaa} & \cite{Villadoro:2006nj} &\cite{Lacour:2007wm} & \cite{Geng:2009ik} \\
 \multicolumn{2}{c}{}& \tiny{\bf BChPT$\times 1/N_c$}  & \tiny{\bf HBChPT;  8+10} & \tiny{\bf HBChPT:  8 only} & \tiny{RBChPT:  8+10} \\
\hline 
$\Lambda p$       & $0.952$ & $-0.048$ & $-0.080$ & $-0.097$ & $-0.031$ \\ 
$\Sigma^-n$       & $0.966$ & $-0.034$ & $-0.024$ & $ 0.008$ & $-0.022$ \\ 
$\Xi^- \Lambda$  & $0.953$ & $-0.047$ & $-0.063$ & $-0.063$ & $-0.029$ \\ 
$\Xi^-\Sigma^0$   & $0.962$ & $-0.038$ & $-0.076$ & $-0.094$ & $-0.030$ \\
\hline\hline
\end{tabular}}
\caption{Corrections to the $\Delta S=1$ vector charges and comparison with previous works.} 
\end{table}

\begin{figure}[]
\begin{center}
\includegraphics[height=5cm,width=8cm,angle=0]{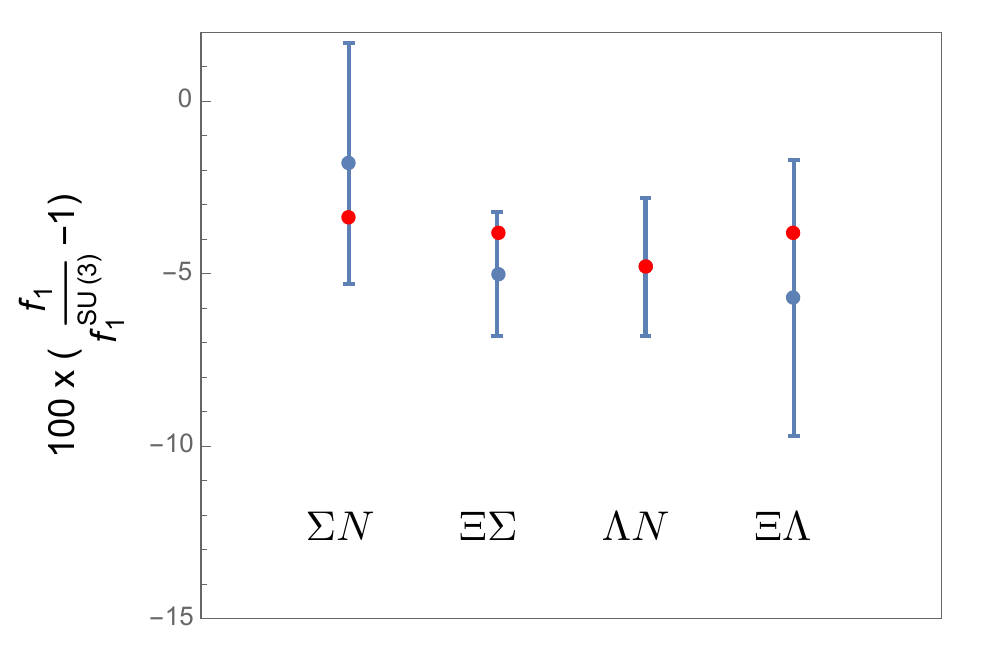}
\end{center}
\caption{One loop corrections to vector current charges. In red the results from the one-loop calculation and in blue the results from a recent LQCD calculation \cite{Shanahan:2015dka}}
\label{fig:2}
\end{figure}

\section{Axial vector currents}
  
  The axial currents play a key role in the implementation of the $1/N_c$ expansion in baryons. As mentioned earlier, their couplings $\ord{N_c}$ to baryons require for consistency of the effective theory that there is a dynamical spin flavor symmetry. To lowest order in \BChPTNc the spatial components of the axial currents are given by $A^{ia}=\gA G^{ia}$. This provides already a first important test of the validity of the spin flavor symmetry, as it locks the relation between the axial couplings of the nucleon and of the $\Delta-N$ transition. Using the $\Delta-N$ transition determined at LO  by the matrix element of the axial current and the resulting  width of the $\Delta$, one obtains that the $\Delta-N$ transition axial coupling is only 2\% smaller than the nucleon's one, confirming the accuracy of the spin flavor symmetry. In the following the one loop corrections are discussed. As observed in LQCD calculations, the quark mass dependencies of the axial couplings  are rather small. Attempts to describe that observation in ordinary BChPT fail precisely because it does not respect $N_c$ power counting. The cancellation of those terms in \BChPTNc~ is quite dramatic, as illustrated by Figure 4 which is the result of a calculation in SU(2) \cite{CalleCordon:2012xz}. This leads naturally to the small non-analytic in quark mass contributions to the axial couplings. In SU(3) with $N_c=3$ at LO the axial couplings of the octet baryons are determined by the two parameters $F$ and $D$,  which in SU(6) are predicted to be $F/D= 2/3$, known to be rather close to the actual physical case.
 \begin{figure}[]
\begin{center}
\includegraphics[height=5cm,width=12cm,angle=0]{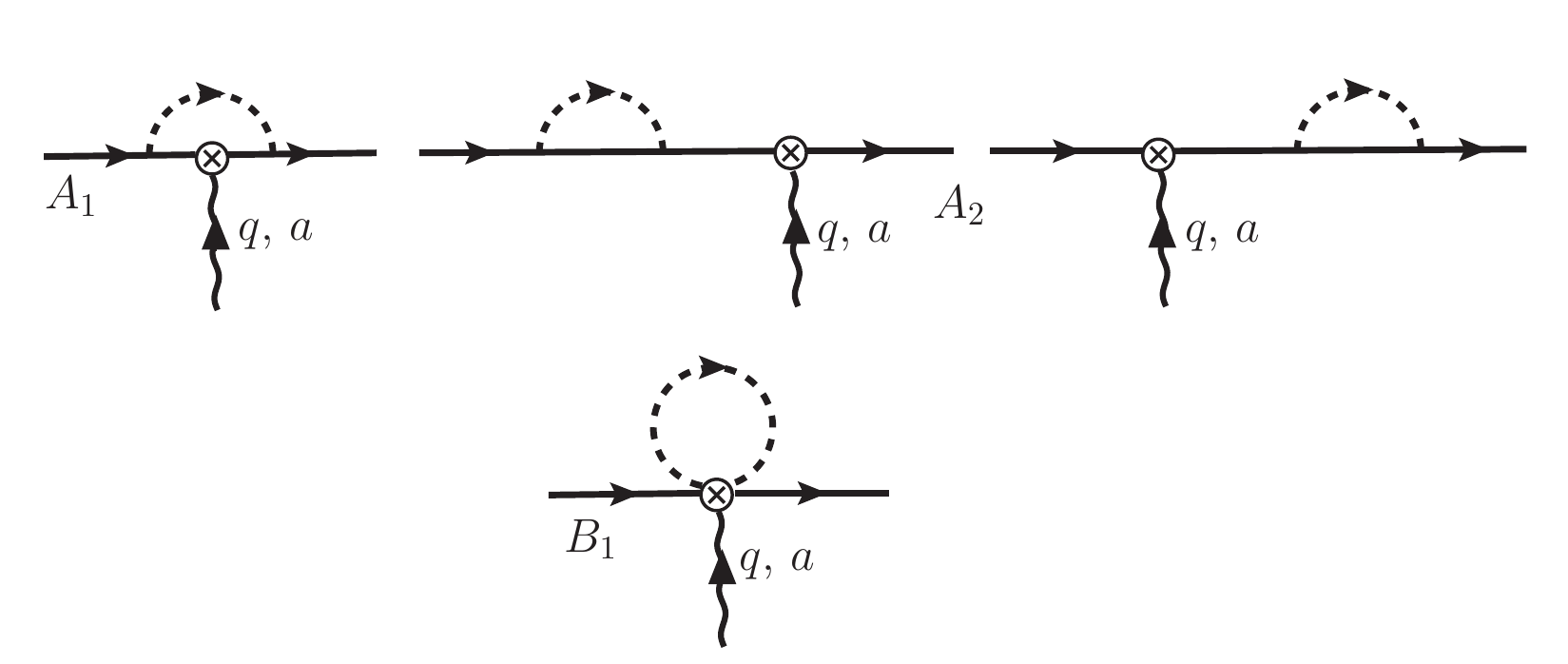}
\end{center}
\caption{One loop corrections to axial vector current couplings.}
\label{fig:1}
\end{figure}
The one loop corrections to the axial couplings are given by the diagrams in Figure 3. Unlike the case of the vector charges, here renormalization is needed. It involves a renormalization of  $\gA$  by a counterterm  $\ord{1/N_c}$, and a number of counterterms: quark mass independent ones that break the spin symmetry, and quark mass dependent ones, a total of nine for general $N_c$, some of which are irrelevant at fixed $N_c=3$ (see \cite{Fernando:2017yqd} for details).   LQCD calculations of axial couplings have a long history, but in the context of SU(3) and including  those of  the decuplet baryons the efforts are rather recent. Ref. \cite{Alexandrou:2016xok} provides results for the  couplings of the two neutral axial currents for both octet and decuplet. In that calculation $m_s$ is kept fixed to approximately its real value,  and $m_u=m_d$ is varied  corresponding to the interval $200 \;\text{ \rm MeV}< M_\pi< 450\; \text{ \rm MeV}$.  Adjusting the definition of axial couplings to those used in \cite{Alexandrou:2016xok},   using \BChPTNc$\; $one fits the LECs corresponding to the mentioned renormalizations. The small quark mass dependency of the axial couplings is naturally described, i.e., with LECs which have natural size magnitude. The LQCD results used here have the well know issue of giving the nucleon's axial couplings  about 10\% smaller than the physical one; this seems to be also the case for the rest of the couplings. One would expect that a corresponding  fudge factor for all couplings can give a realistic adjustment of that issue. Among the important points is that the LO coupling $\gA$ is adjusted by renormalization to be about 20\% smaller that at tree level.  The one loop corrections to the axial couplings are not small for the spin-flavor singlet contributions, i.e., the ones that renormalize $\gA$, while they are small for the rest, including  the quark mass dependent ones. For further tests of the performance of \BChPTNc$\;$  in describing axial couplings it would be good to have LQCD calculations for  the decuplet-octet transition axial couplings, as well as calculations where $m_s$ is also varied, as this would help determine the LECs with more accuracy.
  It is interesting to observe that there are several tests based on combinations of axial couplings which vanish at NNLO tree level receiving  only finite loop corrections \cite{Fernando:2017yqd}.  Such predictions can be used   as tests  of the effective theory vis-\`a-vis  LQCD calculations.  The NNLO results also show that there are no one-loop contributions to the Goldberger-Treiman discrepancies (now generalized to octet and decuplet as well as transition GT discrepancies), with only tree level contributions by two terms (singlet and octet in quark masses) in the $\ord{\xi^3}$ Lagrangian \cite{Fernando:2017yqd}. Finally, recent LQCD calculations of axial form factors \cite{Capitani:2017qpc}  are a strong motivation for extending the analysis of the couplings in \cite{Fernando:2017yqd} to the form factors.
 \begin{figure}[]
\begin{center}
\includegraphics[height=5cm,width=7cm,angle=0]{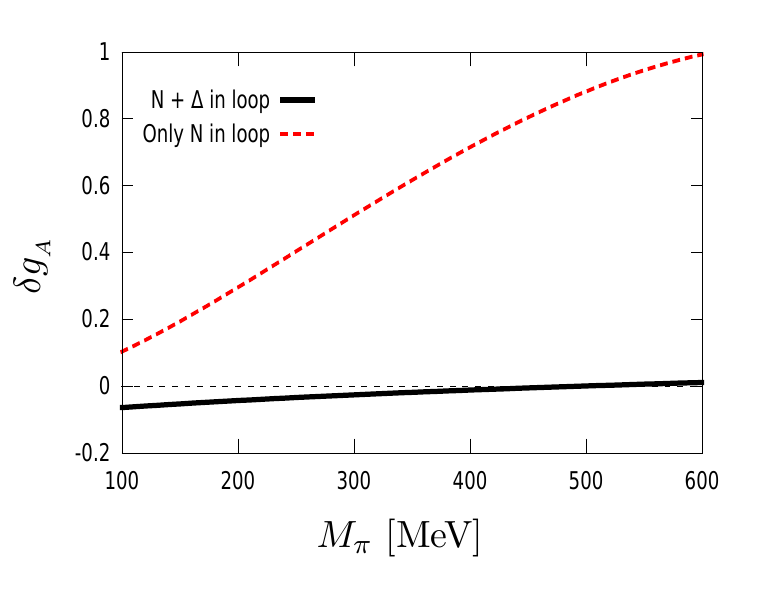}
\end{center}
\caption{Finite loop correction to the axial coupling of the nucleon in $SU(2)$, evaluated at renormalization scale $\mu=m_\rho$, showing  the  dramatic  cancellation effect of including the $\Delta$ contribution. }
\label{fig:3}
\end{figure}

\begin{figure}[]
\begin{center}
\includegraphics[height=4.5cm,width=12cm,angle=0]{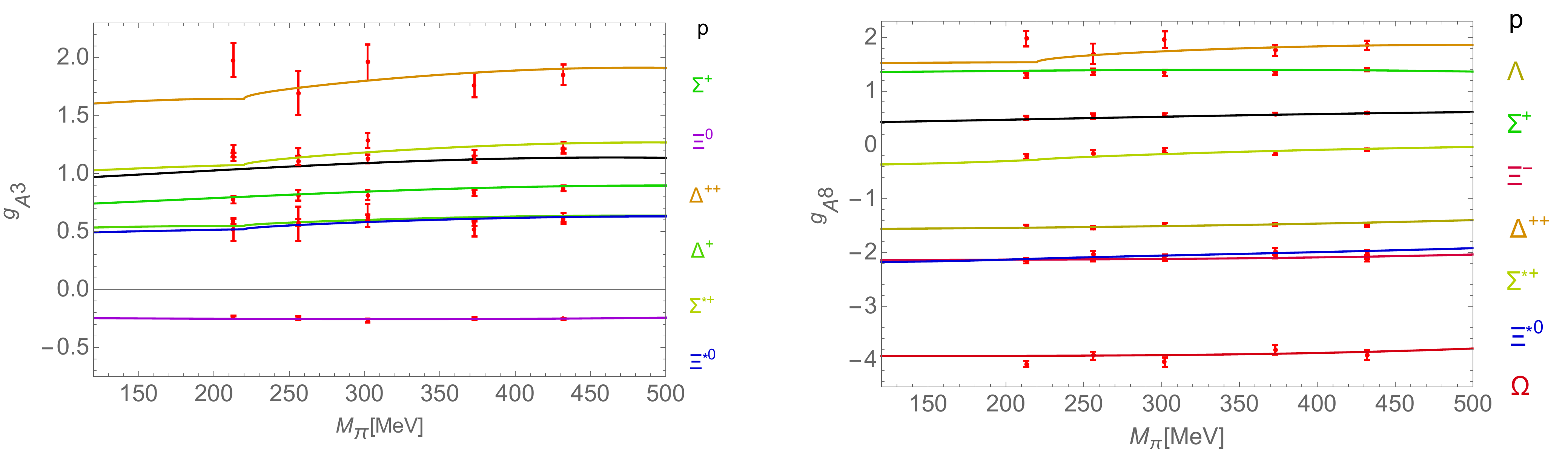}
\end{center}
\caption{Neutral axial currents' couplings  of octet and decuplet baryons. Data  from the LQCD calculation of Ref. \cite{Alexandrou:2016xok}.}
\label{fig:4}
\end{figure}

\section{Summary}

The impact of the $1/N_c$ expansion on BChPT cannot be emphasized strongly enough. It improves the convergence of observables by eliminating $N_c$ power counting violating contributions, leading to vastly improved behavior of the low energy expansion. This has been demonstrated for several observables, namely masses, vector charges and axial couplings as discussed in this note. The case of baryon masses, discussed in detail in \cite{Fernando:2018jrz,FernandoGoityCD18-2}, is of particular interest. It is in the masses where there is a contribution $\ord{ N_c}$, proportional to $(m_u+m_d+m_s)^{\frac 32}$,  which is  a manifestation of the non-commutativity of the chiral and $1/N_c$ expansions; such a contribution is spin-flavor singlet. All contributions which break spin and/or flavor symmetry are however $\ord{1/N_c^n}$ with $n\geq 0$. In particular relations such as Gell-Mann-Okubo and Equal Spacing, which remain valid at generic $N_c$, are corrected by  calculable at NNLO terms $\ord{1/N_c}$, which helps explain the small deviation observed in the physical case. The discussion of the currents presented here also shows very important differences with respect to ordinary BChPT, being the case of the axial currents where the improvements are most significant. Virtually any baryon observable will be affected by the imposition of consistency with the $1/N_c$ expansion, and thus revisiting the most important ones is worthwhile. In addition to current form factors, the applications to $\pi N$ scattering and to Compton scattering are of high interest \cite{Compton,piN}.

\section{Acknowledgements}
 This work was supported by DOE Contract No. DE-AC05-06OR23177 under which JSA operates the Thomas Jefferson National Accelerator Facility, 
and by the  National Science Foundation through grants  PHY-1307413 and PHY-1613951.

\bibliography{Refs}

\end{document}